\documentclass[fleqn,usenatbib]{mnras}

\usepackage{newtxtext,newtxmath}
\usepackage{amsmath}	

\usepackage{amssymb}

\usepackage[T1]{fontenc}

\DeclareRobustCommand{\VAN}[3]{#2}
\let\VANthebibliography\thebibliography
\def\thebibliography{\DeclareRobustCommand{\VAN}[3]{##3}\VANthebibliography}

\usepackage{graphicx}	

\newcommand{\pp}{\texttt{PulsePortraiture}}

\newcommand{\pint}{\texttt{PINT}}
\newcommand{\vela}{\texttt{Vela.jl}}

\newcommand{\orcid}[1]{\href{https://orcid.org/#1}{\textcolor[HTML]{A6CE39}{\includegraphics[height=1.7ex]{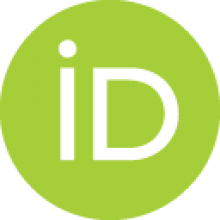}}}}

\title[Revisiting wideband measurements]{Revisiting wideband pulsar timing measurements}

\author[Susobhanan et al.]{
Abhimanyu Susobhanan\textsuperscript{\orcid{0000-0002-2820-0931}}$^{1,2}$\thanks{E-mail: abhimanyu.susobhanan@aei.mpg.de}, 
Avinash Kumar Paladi\textsuperscript{\orcid{0000-0002-8651-9510}}$^{3,1}$,
R{\'e}ka Desmecht\textsuperscript{\orcid{0009-0002-7756-9038}}$^{4,1}$,
Amarnath\textsuperscript{\orcid{0009-0006-0579-3363}}$^{5}$,
\newauthor 
~Manjari Bagchi\textsuperscript{\orcid{0000-0001-8640-8186}}$^{6,7}$,
Manoneeta Chakraborty\textsuperscript{\orcid{0000-0002-9736-9538}}$^{8}$
Shaswata Chowdhury\textsuperscript{\orcid{0000-0001-5701-4014}}$^{6}$,
Suruj Jyoti Das\textsuperscript{\orcid{0000-0002-4144-1799}}$^{9}$,
\newauthor
~Debabrata Deb\textsuperscript{\orcid{0000-0003-4067-5283}}$^{10}$,
Shantanu Desai\textsuperscript{\orcid{0000-0002-8804-650X}}$^{11}$,
Churchil Dwivedi\textsuperscript{\orcid{0000-0002-8804-650X}}$^{12}$,
Himanshu Grover$^{13}$,
Jibin Jose\textsuperscript{\orcid{0009-0009-7652-6758}}$^{8}$,
\newauthor
~Bhal Chandra Joshi\textsuperscript{\orcid{0000-0002-0863-7781}}$^{14,13}$,	
Shubham Kala\textsuperscript{\orcid{0000-0003-2379-0204}}$^{6}$,
Fazal Kareem\textsuperscript{\orcid{0000-0003-2444-838X}}$^{15}$,
Kuldeep Meena\textsuperscript{\orcid{0009-0007-8522-9574}}$^{16}$,
Sushovan Mondal\textsuperscript{\orcid{0000-0002-3657-8256}}$^{6,7}$,
\newauthor
~K Nobleson\textsuperscript{\orcid{0000-0003-2715-4504}}$^{17}$,
B Arul Pandian\textsuperscript{\orcid{0000-0002-0417-6308}}$^{5,18}$,
Kaustubh Rai\textsuperscript{\orcid{0009-0002-2175-7013}}$^{19}$,
Adya Shukla\textsuperscript{\orcid{0009-0005-7058-5539}}$^{13}$,
Manpreet Singh\textsuperscript{\orcid{0009-0001-2715-6641}}$^{20}$,
\newauthor
~Aman Srivastava\textsuperscript{\orcid{0000-0003-3531-7887}}$^{21,11}$,
Mayuresh Surnis\textsuperscript{\orcid{0000-0002-9507-6985}}$^{19}$,
Hemanga Tahbildar\textsuperscript{\orcid{0009-0002-1036-9306}}$^{19}$,
Keitaro Takahashi\textsuperscript{\orcid{0000-0002-3034-5769}}$^{17}$,
\newauthor
~Pratik Tarafdar\textsuperscript{\orcid{0000-0001-6921-4195}}$^{22}$,
Kunjal Vara\textsuperscript{\orcid{0009-0004-5501-1441}}$^{19}$,
Vaishnavi Vyasraj\textsuperscript{\orcid{0009-0008-9261-3870}}$^{11}$,
Zenia Zuraiq\textsuperscript{\orcid{0009-0000-6980-6334}}$^{23}$,
\\
$^{1}$Max-Planck-Institut f{\"u}r Gravitationsphysik (Albert-Einstein-Institut), Liebniz Universit{\"a}t Hannover, Callinstra{\ss}e 38, D-30167 Hannover, Germany\\
$^{2}$School of Physics, Indian Institute of Science Education and Research Thiruvananthapuram, Maruthamala P.O., Thiruvananthapuram 695551, Kerala, India\\
$^{3}$Joint Astronomy Programme, Department of Physics, Indian Institute of Science, C. V. Raman Avenue, Bengaluru 560012, Karnataka, India\\
$^{4}$Vrije Universiteit Brussel, Pleinlaan 2, B-1050 Brussels, Belgium\\
$^{5}$Raman Research Institute, Bengaluru 560080, Karnataka, India\\
$^{6}$The Institute of Mathematical Sciences, C. I. T. Campus, Taramani, Chennai 600113, Tamil Nadu, India\\
$^{7}$Homi Bhabha National Institute, Training School Complex, Anushakti Nagar, Mumbai 400094, Maharashtra, India\\
$^{8}$Department of Astronomy, Astrophysics, and Space Engineering, Indian Institute of Technology Indore, Indore 453552, Madhya Pradesh, India\\
$^{9}$Particle Theory and Cosmology Group, Center for Theoretical Physics of the Universe, Institute for Basic Science, Daejeon 34126, Korea\\
$^{10}$Centre for Space Research, North-West University, Private Bag X6001, Potchefstroom 2520, South Africa\\
$^{11}$Department of Physics, Indian Institute of Technology Hyderabad, Kandi, Telangana 502284, India\\
$^{12}$Astronomy and Astrophysics Division, Physical Research Laboratory, Thaltej Campus, Ahmedabad 380059, Gujarat, India\\
$^{13}$Department of Physics, Indian Institute of Technology Roorkee, Roorkee 247667, Uttarakhand, India\\
$^{14}$National Centre for Radio Astrophysics, Savitribhai Phule Pune University Campus, Pune 411007, Maharashtra, India\\
$^{15}$Max-Planck-Institut für Radioastronomie, 
Auf dem Hügel 69, 53121 Bonn, Germany\\
$^{16}$UM-DAE Centre for Excellence in Basic Sciences, University of Mumbai, Vidyanagari, Mumbai 400098, Maharashtra, India\\
$^{17}$Faculty of Advanced Science and Technology, Kumamoto University, 2-39-1 Kurokami, Kumamoto 860-8555, Japan\\
$^{18}$Department of Physics and Electronics, CHRIST (Deemed to be University), Bengaluru 560029, Karnataka, India\\
$^{19}$Department of Physics, Indian Institute of Science Education and Research Bhopal, Bhopal 462066, Madhya Pradesh, India\\
$^{20}$Department of Physical Sciences, Indian Institute of Science Education and Research Mohali, Sector 81, SAS Nagar, Mohali 140306, Punjab, India\\
$^{21}$Department of Physics, GLA University, Mathura 281406, Uttar Pradesh, India\\
$^{22}$INAF - Osservatorio Astronomico di Cagliari, via della Scienza 5, 09047 Selargius (CA), Italy\\
$^{23}$Department of Physics, Indian Institute of Science, C. V. Raman Avenue, Bengaluru 560012, Karnataka, India
}

\date{Accepted XXX. Received YYY; in original form ZZZ}

\pubyear{\the\year{}}

\begin{document}
\label{firstpage}
\pagerange{\pageref{firstpage}--\pageref{lastpage}}
\maketitle

\begin{abstract}
In the wideband paradigm of pulsar timing, the time of arrival of a pulsar pulse is measured simultaneously with the corresponding dispersion measure from a frequency-resolved integrated pulse profile.
We present a new method for performing wideband measurements that rigorously accounts for measurement noise.
We demonstrate this method using observations of PSR J2124--3358 made as part of the Indian Pulsar Timing Array experiment using the upgraded Giant Metre-wave Radio Telescope, and show that our method produces more realistic measurement uncertainty estimates compared to the existing wideband measurement method.
\end{abstract}




\section{Introduction}
Pulsars are rotating neutron stars whose electromagnetic radiation appears as periodic pulses to a terrestrial observer \citep{LorimerKramer2012}.
They are characterized by high rotational stability, acting as accurate celestial clocks.
Pulsar timing is the technique of tracking a pulsar's rotation by measuring the times of arrival (TOAs) of its pulses \citep{HobbsEdwardsManchester2006}.
Pulsar timing is one of the most precise techniques in astronomy with a wide range of applications, from spacecraft navigation \citep{DengHobbs+2013} to gravitational wave detection \citep{AgazieAntoniadis+2024}.

Radio signals emitted by a pulsar are dispersed as they propagate through the ionized interstellar medium (ISM), introducing a delay that is proportional to the electron column density along the line of sight and inversely proportional to the square of the radio frequency \citep{BackerHellings1986}.
The electron column density towards a pulsar, known as the dispersion measure (DM), varies stochastically as a function of time due to the dynamic nature of the ISM and the relative motion of the pulsar with respect to the Earth.
In addition to the pulsar rotation and interstellar dispersion, the TOAs measured at a terrestrial observatory are influenced by several astrophysical phenomena, including geometric and relativistic delays due to the orbital motion of the pulsar and the Earth, passing gravitational waves, and instrumental delays and noise \citep{HobbsEdwardsManchester2006,EdwardsHobbsManchester2006,DamourDeruelle1986,EstabrookWahlquist1975}.
Pulsar timing involves characterizing all of these effects to construct a predictive model for the TOAs.

A pulsar observation using a radio telescope generates an observing frequency-resolved radio light curve.
This light curve is folded using the known rotational frequency of the pulsar to obtain an integrated pulse profile, which gives the pulsar intensity as a function of the pulse phase and the observing frequency.
In conventional pulsar timing, also known as narrowband timing, a frequency-resolved integrated pulse profile spanning multiple frequency sub-bands covering the observed band-pass is produced, and a TOA is measured from each sub-band independently by cross-correlating it against a similarly frequency-resolved template profile \citep{Taylor1992}.
On the other hand, in the wideband timing paradigm, a single TOA along with a DM is measured from a frequency-resolved two-dimensional integrated pulse profile, also known as a portrait, by cross-correlating it against a two-dimensional template \citep{PennucciDemorestRansom2014,LiuDesvignes+2014}.
Wideband timing can also be performed on combined datasets consisting of simultaneous multi-band observations \citep{PaladiDwivedi+2024}.
The two-dimensional templates for wideband timing, also known as template portraits, are usually obtained using a principal component analysis-based method, described in \citet{Pennucci2019}.
In general, a wideband timing measurement produces the TOA and DM estimates, their uncertainties, and their covariance, although it turns out that the measurement covariance can be made to vanish by adjusting the fiducial frequency at which the measurement is made; see Section \ref{sec:wb-measurement}.
Consequently, wideband timing and noise analysis are done by incorporating both the TOA and the DM measurements \citep{AlamArzoumanian+2021,SusobhananVan_Haasteren2025,Susobhanan2025b}.
{Algorithms for generating template portraits and for performing wideband TOA and DM measurements are implemented in the \pp{} package \citep{PennucciDemorestRansom2014,Pennucci2019}, wideband timing methods are implemented in \texttt{tempo} \citep{NiceDemorest+2015} and \pint{} \citep{LuoRansom+2021,SusobhananKaplan+2024}, and Bayesian wideband noise analysis is implemented in \texttt{ENTERPRISE} \citep{EllisVallisneri+2020} and \vela{} \citep{Susobhanan2025b}.}
Wideband timing provides two main advantages over narrowband timing --- reduced data volumes and better handling of frequency-dependent profile variability \citep{HankinsRickett1986}.

In this work, we present an improved method for wideband TOA and DM measurements, which provides better handling of the noise present in the observed portrait, thereby producing more realistic estimates of the measurement uncertainty.
This paper is arranged as follows.
In Section \ref{sec:wb-measurement}, we derive our new method for wideband measurements, starting from first principles. 
In Section \ref{sec:sim-ex}, we demonstrate our method using a simulated example profile.
In Section \ref{sec:inpta-ex}, we demonstrate our methods using data obtained as part of the Indian Pulsar Timing Array (InPTA) experiment \citep{JoshiArumugasamy+2018,TarafdarNobleson+2022,RanaTarafdar+2025}.
Finally, we summarize our results in Section \ref{sec:summary}.

\section{Revisiting wideband measurements}
\label{sec:wb-measurement}

We begin by writing down a model for the observed total intensity portrait as a function of the pulse phase $\phi$ and the observing frequency $\nu$ following \citet{PennucciDemorestRansom2014}.
\begin{equation}
    P(\phi, \nu) = b(\nu) + a(\nu)T(\nu, \phi - \varphi (\nu)) + n(\phi, \nu)\,,
    \label{eq:portrait}
\end{equation}
where $P(\phi,\nu)$ is the observed portrait, $T(\phi,\nu)$ is the template portrait, $b(\nu)$ is the baseline total intensity, $a(\nu)$ is the amplitude of the signal at each frequency, $\varphi(\nu)$ is a frequency-dependent phase shift, and $n(\phi,\nu)$ is a white noise process.
The frequency dependence of $a(\nu)$ and $b(\nu)$ arises because of the pulsar's intrinsic emission spectrum \citep{MaronKijak+2000}, interstellar scintillation \citep{Rickett1970}, and the shape of the telescope bandpass.
The phase shift $\varphi(\nu)$ is given by
\begin{equation}
    \varphi(\nu) = \varphi_0 - \kappa D \bar{F} \left( \nu^{-2} - \nu_\text{ref}^{-2}\right),
    \label{eq:varphi}
\end{equation}
where $\varphi_0$ is the total phase shift due to achromatic effects, $\kappa$ is the DM constant, $D$ is the DM at the observing epoch, $\bar{F}$ is the topocentric frequency of the pulsar at the observing epoch, and $\nu_\text{ref}$ is an arbitrary fiducial observing frequency.
The phase shift $\varphi_0$ is related to the wideband TOA $t_\text{arr}$ as
\begin{equation}
    t_\text{arr} = t_\text{obs} + \bar{F}^{-1} \varphi_0\,,
\end{equation}
where $t_\text{obs}$ is the fiducial observation time, usually taken to be at the middle of the observation span.
The noise term $n(\phi,\nu)$ arises due to radiometer noise, pulse jitter, radio frequency interference (RFI), etc.

In practice, the observed portrait is discrete, and equation \eqref{eq:portrait} can be written in discrete form as
\begin{equation}
    P_{\alpha j} = b_\alpha + a_\alpha T(\nu_\alpha, \phi_j - \varphi_\alpha) + n_{\alpha j}\,,
\label{eq:portrait_discrete}
\end{equation}
where $\alpha \in \{1, ..., N_\text{chan}\}$ represents the observing frequency channel, $N_\text{chan}$ is the number of frequency channels, $j \in \{1, ..., N_\text{bin}\}$ represents the pulse phase bin, $N_\text{bin}$ is the number of phase bins, and $\varphi_\alpha = \varphi_0 - \kappa D \bar{F} \left( \nu_\alpha^{-2} - \nu_\text{ref}^{-2}\right)$.
Following \citet{Taylor1992} and \citet{PennucciDemorestRansom2014}, we compute the discrete Fourier transform of equation \eqref{eq:portrait_discrete} along the phase index.
Applying the Fourier shift theorem and discarding the zero-frequency baseline term, we obtain
\begin{equation}
    \tilde{P}_{\alpha k} = a_\alpha \tilde{T}_{\alpha k} \exp[-2 \pi i k \varphi_\alpha] + \tilde{n}_{\alpha k}\,,
\end{equation}
where tilde ($\,\tilde{}\,$) represents the discrete Fourier transform, and  $k \in \{1, ..., N_\text{bin}\}$.
Assuming $\tilde{n}_{\alpha k}$ to be an uncorrelated \textit{complex} Gaussian process with an observing frequency-dependent variance $\sigma_\alpha^2$, we can write down a Fourier-domain log-likelihood function for the observed portrait:
\begin{align}
\ln L &= \sum_{\alpha=1}^{N_{\text{chan}}}\sum_{k=1}^{N_{\text{bin}}/2}\left[\frac{-\left|\tilde{P}_{\alpha k}-a_{\alpha}\tilde{T}_{\alpha k}\exp[-2\pi ik\varphi_{\alpha}]\right|^{2}}{\sigma_{\alpha}^{2}}-\ln[\pi\sigma_{\alpha}^{2}]\right]\nonumber
\\&=-\sum_{\alpha=1}^{N_{\text{chan}}}\left[\sigma_{\alpha}^{-2}\left(U_{\alpha}+a_{\alpha}^{2}V_{\alpha}-2a_{\alpha}W_{\alpha}\right)+
\frac{N_{\text{bin}}}{2}\ln[\pi\sigma_{\alpha}^2]\right]\,.
\label{eq:lnlike}
\end{align}
We only consider Fourier frequencies up to $N_\text{bin}/2$ in the summation because both $P_{\alpha j}$ and $T_{\alpha j}$ are real-valued.
The quantities $U_\alpha$, $V_\alpha$, and $W_\alpha$ are real-valued and are defined as
\begin{subequations}
\begin{align}
U_{\alpha}&=\sum_{k=1}^{N_{\text{bin}}/2}\left|\tilde{P}_{\alpha k}\right|^{2}\,,\\
V_{\alpha}&=\sum_{k=1}^{N_{\text{bin}}/2}\left|\tilde{T}_{\alpha k}\right|^{2}\,,\\
W_{\alpha}&=\sum_{k=1}^{N_{\text{bin}}/2}\Re\left[\tilde{P}_{\alpha k}\tilde{T}_{\alpha k}^{*}\exp[2\pi ik\varphi_{\alpha}]\right]\,.
\label{eq:W}
\end{align}
\end{subequations}
Here, $W_\alpha$ is a function of $\varphi_0$, $D$, and $\nu_\text{ref}$, whereas $U_\alpha$ and $V_\alpha$ only depend on the data and the template portraits, respectively.
 
$\ln L$ is a function of the parameters of interest $\varphi_0$ and $D$, the nuisance parameters $a_\alpha$ and $\sigma_\alpha$, and the arbitrary constant $\nu_\text{ref}$.
\citet{PennucciDemorestRansom2014} removes the $a_\alpha$ dependence by maximizing equation \eqref{eq:lnlike} over those parameters, and fixes $\nu_\text{ref}$ by demanding that the measurement covariance between $\varphi_0$ and $D$ vanish.
Multiple methods have been employed to estimate the noise standard deviation $\sigma_\alpha$.
This includes using the variance within a window around the profile's minimum, assuming it lies in the off-pulse region, or estimating the variance based on the highest harmonics in the discrete Fourier transform $\tilde{P}_{\alpha k}$ \citep{PennucciDemorestRansom2014}. 
The latter method is implemented in \pp.
Unfortunately, these approaches do not always provide reliable noise estimates: for example, the former method fails for pulsars with large duty cycles 
and the latter fails when the profile signal-to-noise ratio is sufficiently high, making even the higher harmonics dominated by the signal.
The latter failure mode occurs when (a) the pulsar is bright, (b) the telescope has a large collecting area, and/or (c) the pulsar is observed for a long duration.

To overcome this limitation, we explore a Bayesian approach to do away with the nuisance parameters.
The Bayes' theorem in this context can be written as
\begin{align}
\mathfrak{p}[\varphi_0,D,\boldsymbol{a},\boldsymbol{\sigma}|P,T] =& L[P|\varphi_0,D,\boldsymbol{a},\boldsymbol{\sigma};T] \nonumber\\&\Pi[\varphi_0]\,\Pi[D]\, \Pi[\boldsymbol{a}]\,\Pi[\boldsymbol{\sigma}] / Z[P|T]\,,
\label{eq:post-full}
\end{align}
where $\mathfrak{p}$ denotes the posterior distribution, the likelihood $L$ is given by equation \eqref{eq:lnlike}, $\Pi$ denotes the prior distributions, $Z$ is the Bayesian evidence, and $\boldsymbol{a}$ and $\boldsymbol{\sigma}$ represent the collection of $a_\alpha$ and $\sigma_\alpha$ parameters respectively.
We can marginalize equation \eqref{eq:post-full} analytically over $a_\alpha$ and $\sigma_\alpha$ by imposing the appropriate conjugate priors on those parameters.
{We choose the Jeffreys' priors \citep{Jeffreys1946} for these parameters, which correspond to an improper uniform prior for $a_\alpha$ and an improper log-uniform prior for $\sigma_\alpha$; i.e.,
\begin{subequations}
\begin{align}
\Pi[a_\alpha]&\propto 1\,,\\
\Pi[\sigma_\alpha]&\propto \sigma_\alpha^{-1}\,.
\end{align}
\end{subequations}}

Marginalizing equation \eqref{eq:post-full} over the $a_\alpha$ and $\sigma_\alpha$ using the above-mentioned priors, we obtain
\begin{align}
\mathfrak{p}[\varphi_{0},D|P,T]=\Pi[\varphi_{0}]\,\Pi[D]\,\Lambda[P|\varphi_{0},D]\,,
\label{eq:post-marg}
\end{align}
where the marginalized log-likelihood can be written, up to an additive constant, as
\begin{align}
    \ln \Lambda =\frac{-(N_{\text{bin}}-1)}{2}\sum_{\alpha}\ln\left(U_{\alpha}-\frac{W_{\alpha}^{2}}{V_{\alpha}}\right)\,.
    \label{eq:loglambda}
\end{align}
A detailed derivation of equation \eqref{eq:loglambda} is provided in Appendix \ref{sec:loglambda}.

We note in passing that an alternative to marginalization is the profile likelihood approach \citep{HeroldFerreiraHeinrich2025}, where the likelihood function is \textit{maximized} over the nuisance parameters.
This approach yields an expression that is very similar to equation \eqref{eq:loglambda}:
\begin{equation}
    \ln\hat{L}	= \frac{-N_{\text{bin}}}{2}\sum_{\alpha}\ln\left(U_{\alpha}-\frac{W_{\alpha}^{2}}{V_{\alpha}}\right)\,;
    \label{eq:loglhat1}
\end{equation}
see Appendix \ref{sec:maxlike} for a detailed derivation.
A careful inspection of these expressions reveals that both equations \eqref{eq:loglambda} and \eqref{eq:loglhat1} provide the same maximum-likelihood estimate, but equation \eqref{eq:loglambda} provides slightly broader and more conservative measurement uncertainty estimates.
Note that we use equation \eqref{eq:loglambda} exclusively in the remainder of this paper, and equation \eqref{eq:loglhat1} is shown only for comparison.

Since the fiducial frequency $\nu_\text{ref}$ is arbitrary, it is usually chosen such that the measurement covariance between $\varphi_0$ and $D$ vanishes, so that they can be treated as independent measurements, in turn simplifying and reducing the computational cost of wideband timing and noise analysis \citep{AlamArzoumanian+2021,SusobhananVan_Haasteren2025}.
This is achieved in \citet{PennucciDemorestRansom2014} using a Hessian-based analytic estimate of the measurement covariance, by setting the second derivative $\frac{\partial^2 \ln L}{\partial \varphi_0 \partial D}$ to zero.
This choice of the fiducial frequency for our noise-marginalized likelihood, denoted $\bar\nu_\text{ref}$, is derived in Appendix \ref{sec:nurefbar}.
It can be seen from Appendix \ref{sec:nurefbar} that (a) the $\bar\nu_\text{ref}$ depends on the observed portrait on each epoch, and therefore is different for each epoch, and (b) the $\bar\nu_\text{ref}$ estimate is derived assuming a Gaussian posterior distribution, and may not be accurate when the posterior is significantly non-Gaussian, e.g., in the presence of loud RFI.
The measurement uncertainties in $\varphi_0$ and $D$ can also be estimated with the help of the Hessian, and these computations are also given in Appendix \ref{sec:nurefbar}.

\section{A simulated example}
\label{sec:sim-ex}

\begin{figure}
   \centering
   \includegraphics[width=0.46\textwidth]{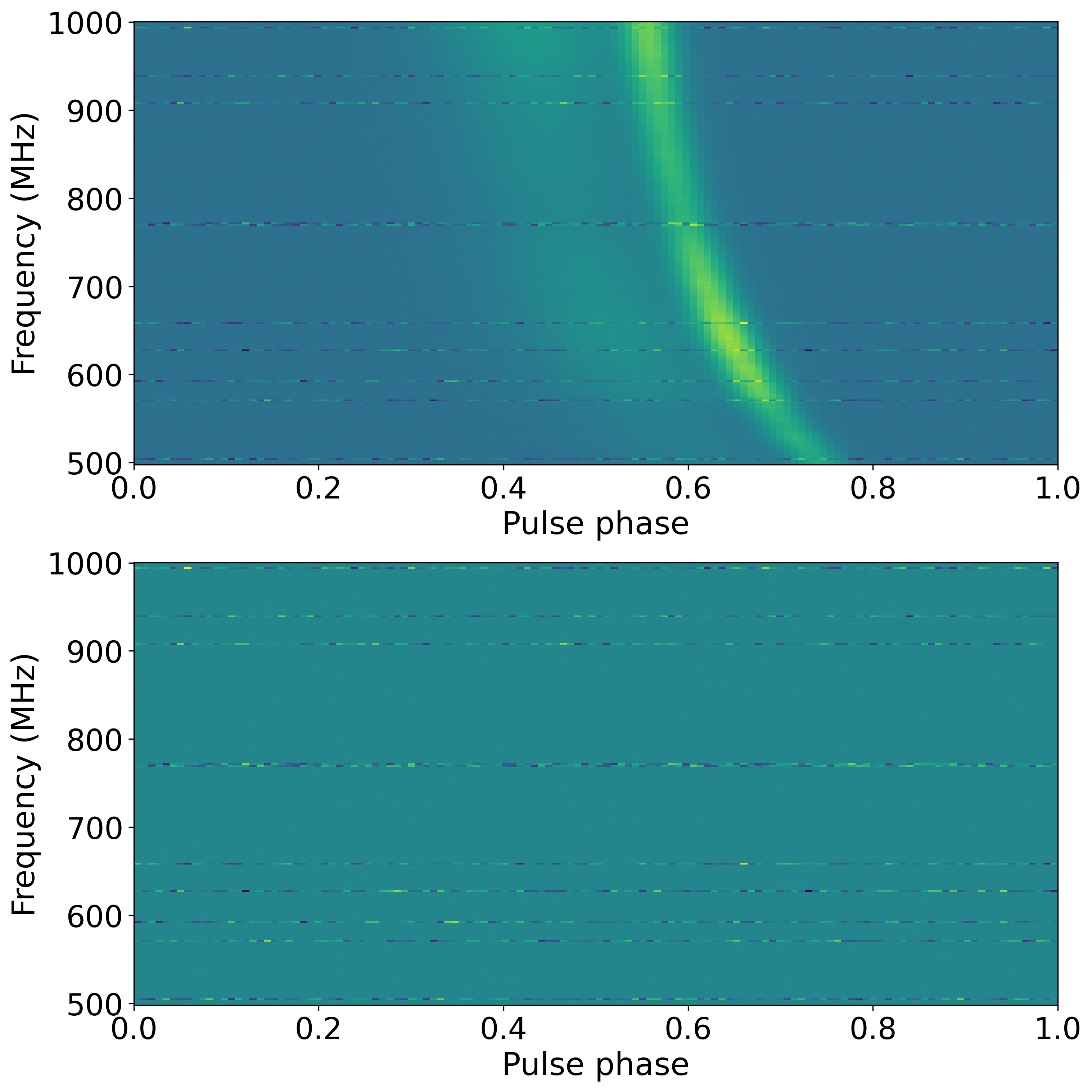}
   \caption{The top plot shows the simulated pulse portrait based on the analytic two-peaked template given in equation \eqref{eq:templ-sim}. The template is weighted by a dynamic spectrum emulating scintillation. It also contains white noise in each frequency channel.
   Ten randomly selected channels have noise injections with a significantly higher amplitude, simulating RFI, and these are visible in the plot as horizontal stripes. 
   The bottom plot shows the post-fit residuals of the portrait, which are determined by subtracting the best-fitting model from the simulated data.
   The residuals show only noise, indicating a successful fit. 
    }
   \label{fig:dataprof}
\end{figure}

We now present an example to illustrate our method using a simulated portrait. 
We assume a frequency-dependent two-peaked template portrait of the form
\begin{equation}
    T(\phi,\nu)=\frac{3}{5}\text{vm}\left[\phi;\pi,60\right]+\frac{2}{5}\left(\frac{\nu}{400\text{\,MHz}}\right)\text{vm}\left[\phi;\frac{3\pi}{4},5\right]\,,
    \label{eq:templ-sim}
\end{equation}
where $\text{vm}[\phi;\mu,\chi]$ is the von Mises function given by
\begin{align*}
    \text{vm}[\phi;\mu,\chi] = \frac{\exp(\chi \cos(\phi - \mu))}{ 2\pi I_0(\chi)}\,.
\end{align*}
We simulate an observed portrait using equation \eqref{eq:portrait_discrete} with $b(\nu)=0.2$, $\varphi_0=0.1$, $D=0.15$ pc/cm$^3$, $\nu_{\text{ref}}=750$ MHz, $\bar{F}=100$ Hz, $N_\text{bin}=128$, and $N_\text{chan}=258$ for observing frequencies between 500--1000 MHz. 
{We simulate frequency-dependent amplitude modulations due to interstellar scintillation using the \texttt{SCINTOOLS} package \citep{ReardonColes+2020}\footnote{We set the Max Born scattering strength parameter ($m_b^2$) to be 15 and the structure function exponent to be 5/3.}}.
We inject white noise realizations with a standard deviation $\sigma_\alpha=0.025$ in every frequency channel, except for 10 randomly selected channels where we use $\sigma_\alpha=0.25$ to simulate RFI.
The resulting simulated portrait is shown in {the top panel of} Figure  \ref{fig:dataprof}. 

\begin{figure}
   \centering
\includegraphics[width=0.45\textwidth]{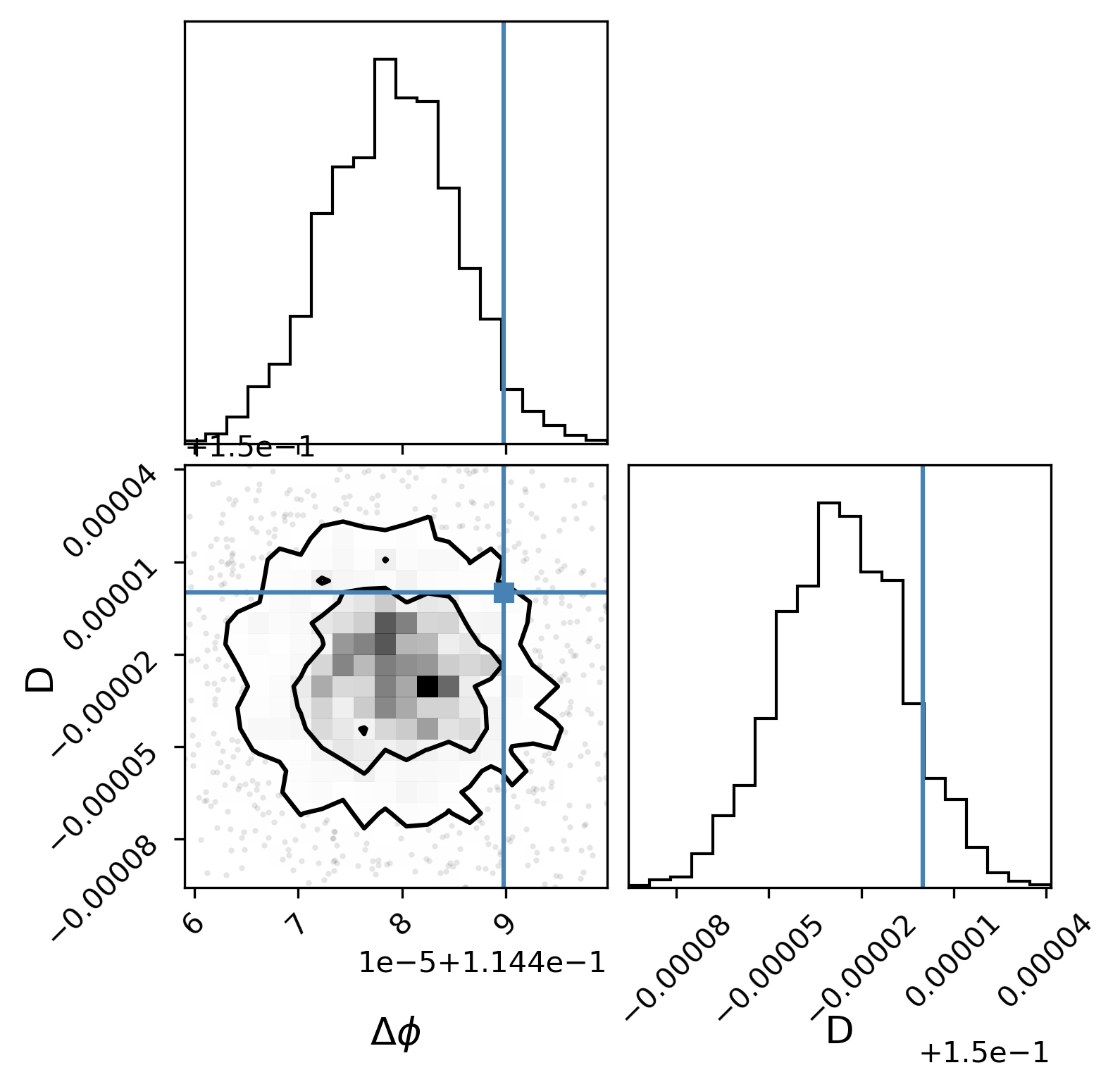}
   \caption{The posterior distribution obtained from the simulated portrait described in Section \ref{sec:sim-ex}.
   This analysis was done using a fiducial frequency value such that the measurement covariance between the two parameters vanishes (i.e., $\bar{\nu}_\text{ref}$), as described in Appendix \ref{sec:nurefbar}.
   The blue lines represent the injected values shifted according to the $\bar{\nu}_\text{ref}$.
   This corresponds to $\varphi_{0\text{;inj}}-\kappa DF\left(\nu_{\text{ref;inj}}^{2}-\bar{\nu}_{\text{ref}}^{2}\right)$ for the $\varphi_0$ measurement whereas the $D$ measurement is expected to be consistent with $D_\text{inj}$, where the label `inj' represents the injected value.
   The simulated portrait and the post-fit residuals are plotted in Figure \ref{fig:dataprof}.
   The estimated parameters are consistent with the injected values up to a shift described above.
   The contours represent  68\% and 95\% credible intervals.
   }
   \label{fig:cornerplot}
\end{figure}

We estimate $\varphi_0$ and $D$ from this simulated template using the amplitude and noise-marginalized likelihood function {given in} equation \eqref{eq:loglambda}.
Assuming broad uniform priors for $\varphi_0$ and $D$, we draw samples from the posterior distribution (equation \ref{eq:post-marg}) using the MultiNest nested sampling algorithm \citep{FerozHobsonBridges2009}, implemented in the \texttt{nestle} package \citep{Barbary2021}.
The posterior distribution $p[\varphi_0,D|P,T]$ at a fiducial frequency $\bar{\nu}_\text{ref}$ estimated based on Appendix \ref{sec:nurefbar} is plotted in Figure \ref{fig:cornerplot}.
The portrait residuals, obtained by subtracting the best-fit model from the data portrait in this case, are plotted in the bottom panel of Figure \ref{fig:dataprof}.
We find that the injected $\varphi_0$ and $D$ values are reliably estimated, and that the residuals in each frequency channel are white.

\begin{figure}
   \centering
   \includegraphics[width=0.4\textwidth]{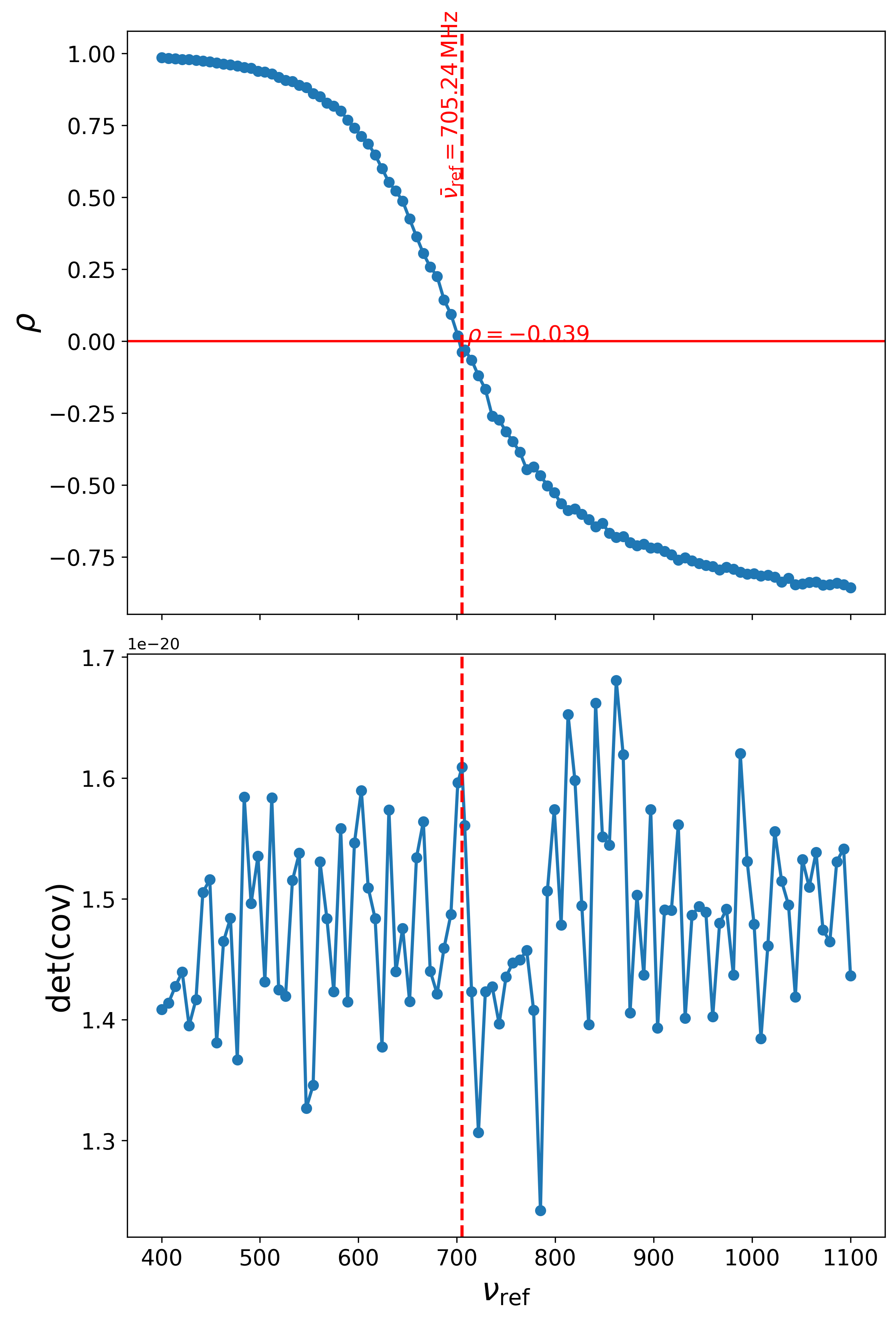}
   \caption{The top plot shows the measurement correlation $\rho$ between $\varphi_0$ and $D$ as a function of $\nu_\text{ref}$.
   The $\bar{\nu}_\text{ref}$ value computed based on Appendix \ref{sec:nurefbar} is indicated using a vertical dotted line, and the corresponding correlation estimated from posterior samples is indicated using a horizontal red line.
   The correlation at $\bar{\nu}_\text{ref}$ is very close to zero.
   The bottom plot shows the determinant of the covariance matrix between $\varphi_0$ and $D$ as a function of ${\nu}_\text{ref}$.
   This determinant is a measure of the total measurement uncertainty, and does not show any significant variation as a function of ${\nu}_\text{ref}$ except for random scatter.
   The scatter seen in the two plots is due to the numerical error from using a finite number of posterior samples.}
   \label{fig:correlations}
\end{figure}

We now proceed to ascertain the accuracy of the $\bar{\nu}_\text{ref}$ obtained in Appendix \ref{sec:nurefbar}.
For this purpose, we estimate $\varphi_0$ and $D$ using 101 uniformly separated values of $\nu_{\text{ref}}$ in the 400--1100 MHz range using the noise-marginalized likelihood of equation \eqref{eq:loglambda} using nested sampling.
We find that the estimated $\phi_0$ and $D$ values are consistent with the injected values for every value of $\nu_{\text{ref}}$, as expected\footnote{The measured $\phi_0$ value depends on $\nu_\text{ref}$. We deem the measured $\phi_0$ value to be consistent with the injected value if $\left|\phi_{0}^\text{meas}-\phi_{0}^\text{inj}-\kappa D F({\nu}_\text{ref}^\text{meas}-{\nu}_\text{ref}^\text{inj})\right|$ is less than the measurement uncertainty of $\phi_0$, where the superscripts `meas' and `inj' indicate the measured and injected values. On the other hand, the measured $D$ value should be independent of $\nu_\text{ref}$.}.
We plot the Pearson correlation coefficient between $\varphi_0$ and $D$ as a function of $\nu_{\text{ref}}$ in Figure \ref{fig:correlations}.
We find that the correlation becomes approximately zero at $\nu_{\text{ref}}=\bar\nu_{\text{ref}}$, although a tiny correlation remains even at this $\nu_{\text{ref}}$ value\footnote{A similar observation was also made by \citet{PennucciDemorestRansom2014}}.
We also plot the determinant of the covariance matrix, which is a measure of the total uncertainty in the wideband measurement, as a function of $\nu_{\text{ref}}$, in Figure \ref{fig:correlations}.
We find that this determinant value remains independent of $\nu_{\text{ref}}$ up to numerical precision, indicating that the overall measurement precision is not affected by the choice of $\nu_{\text{ref}}$.


\section{Demonstration using \texorpdfstring{I\lowercase{n}PTA}{InPTA} observations of PSR J2124--3358}
\label{sec:inpta-ex}

We now proceed to demonstrate our new wideband measurement method using frequency-resolved integrated pulse profiles of PSR J2124--3338 and compare it against the wideband measurement method of \citet{PennucciDemorestRansom2014}. 
PSR J2124--3358 is a bright, nearby, and isolated millisecond pulsar with a rotational period $P_s \sim$ 4.93 ms, located at a distance of around 400 pc. 
It was discovered using the Parkes 64-m Radio Telescope in the Parkes 436 MHz survey of the southern sky \citep{BailesJohnston+1997}. 
Its stable, precisely timed pulses make it an important target for pulsar timing array (PTA) experiments \citep[e.g. ][]{RanaTarafdar+2025}.

The data used in this section include integrated pulse profiles obtained from observations done using the upgraded Giant Metre-wave Radio Telescope \citep[uGMRT:][]{GuptaAjithkumar+2017} as part of the InPTA experiment \citep{JoshiArumugasamy+2018} during observation cycles 37--48 (MJD 58781--60896) in the 300--500 MHz band (Band 3).
TOAs derived from subsets of this data were included in the first and second data releases of the InPTA \citep{TarafdarNobleson+2022,RanaTarafdar+2025}; however, the epochs 60426--60896 have not been published elsewhere.
The observation and data reduction procedures used to obtain this dataset are described in \citet{SusobhananMaan+2021},  \citet{TarafdarNobleson+2022}, and \citet{RanaTarafdar+2025}. 
We have excluded observing epochs significantly affected by RFI from this analysis.

The wideband measurement technique described in \cite{PennucciDemorestRansom2014} and implemented in the $\pp$ package is henceforth referred to as the standard timing technique.
In the standard technique, the noise parameters $\sigma_\alpha$ are estimated by considering the mean of the last 25\%\footnote{This fraction (25\%) can be adjusted while running \pp{}.
Unfortunately, there is no rigorous procedure to determine the optimal value of this fraction, so we have persisted with the default value used in \pp{}. Further, this value has been used in previous InPTA wideband analyses \citep{NoblesonAgarwal+2022,TarafdarNobleson+2022}.} of the Fourier bins in the power spectrum computed from $\tilde{P}_{\alpha k}$. 
We have implemented the amplitude and noise-marginalized likelihood given in equation \eqref{eq:loglambda} in $\pp$, and this is referred to as MLAN\footnote{Marginalized Likelihood over Amplitude and Noise} hereafter.

\begin{figure}
   \centering
   \includegraphics[width=0.48\textwidth]{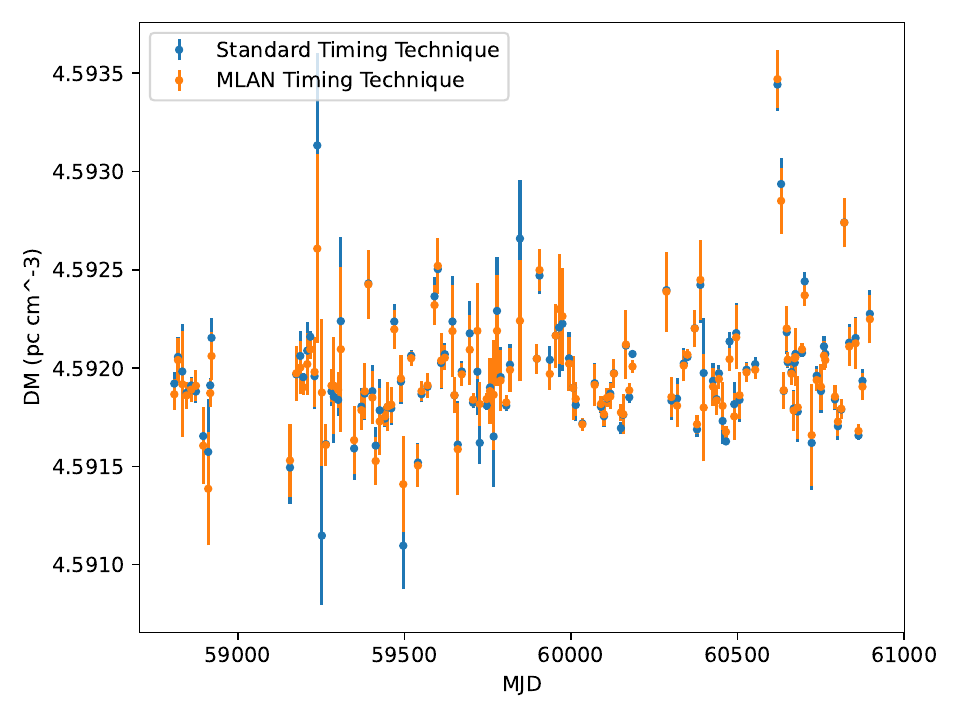}
   \caption{DM time series for PSR J2124--3358 obtained using the standard and MLAN measurement methods.
   The measurements are consistent but not identical, except for a few epochs where there is a significant difference between the two methods.}
   \label{fig:DMtimeseries}
\end{figure}

We applied these two methods to the frequency-resolved integrated pulse profiles of PSR J2124--3338 to measure the wideband TOAs and DMs.
The wideband DM time series obtained using the two methods are shown in Figure \ref{fig:DMtimeseries}. 
We find that the DM estimates obtained using the two methods are not always identical, and there are a few epochs where the measurements are inconsistent.
Nevertheless, the DM measurements are consistent on the majority of epochs.

\begin{figure*}
   \centering
   \includegraphics[width=1\textwidth]{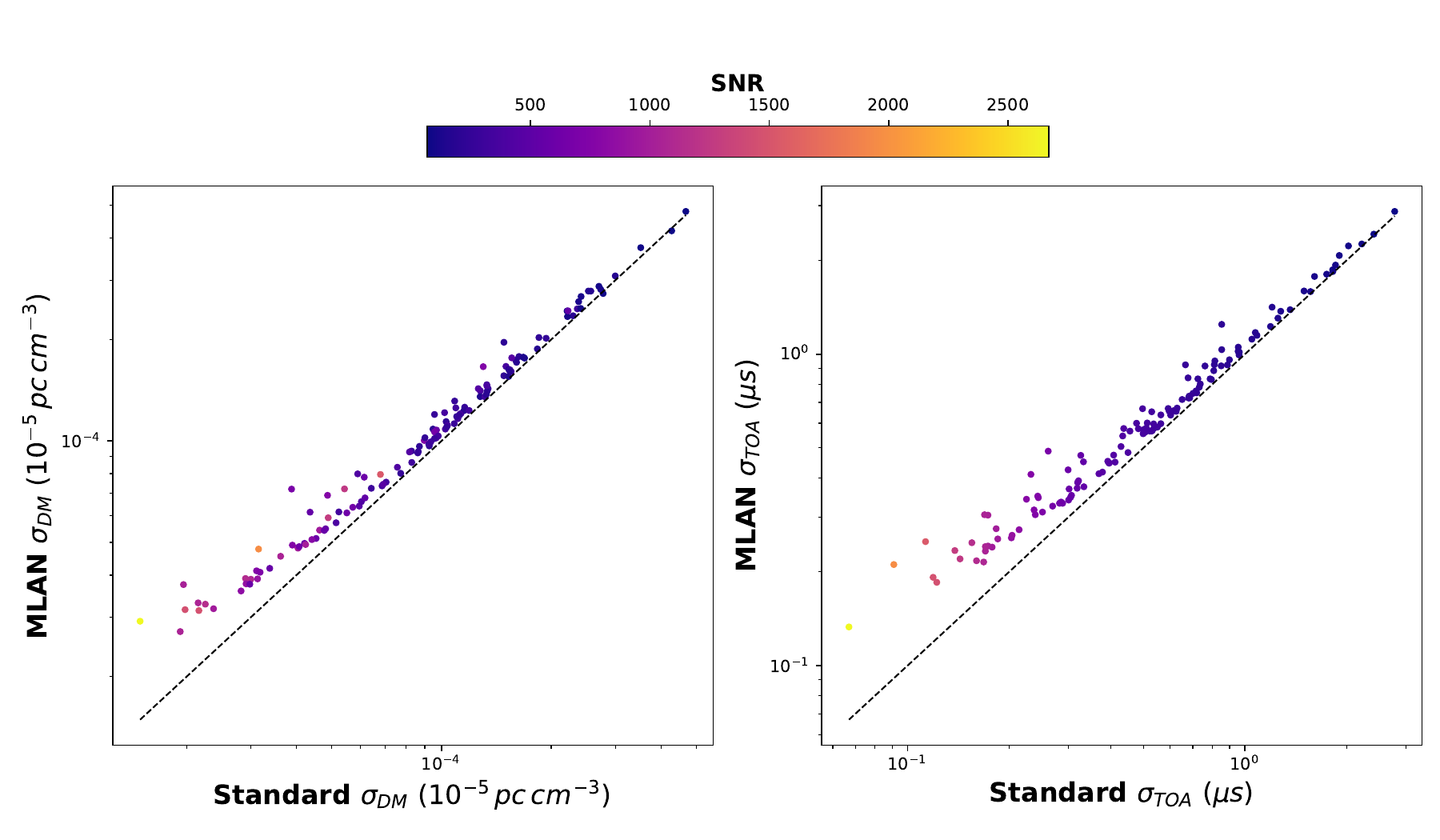}
   \caption{Comparison of TOA and DM measurement uncertainties obtained using the standard and MLAN methods.
   The S/N at each epoch is represented as the color of each point.
   The MLAN uncertainties are consistently higher than their standard method counterparts.
   This is most prominent in high-S/N epochs.}
   \label{fig:errcomp}
\end{figure*}

A comparison of the TOA and DM measurement uncertainties obtained using the two methods is shown in Figure \ref{fig:errcomp}.
We find that the measurement uncertainties obtained with the MLAN method are consistently greater than those obtained with the standard method.
In particular, the discrepancy in uncertainties is larger for high-S/N epochs.
We argue that this is because the $\sigma_\alpha$ estimates used in the standard method, computed from higher harmonics in the power spectrum derived from $\tilde{P}_{\alpha k}$, are inaccurate, especially when the S/N is high, where even the higher Fourier bins in the power spectrum will be dominated by signal.
On the other hand, the MLAN method provides a rigorous treatment of the noise in each frequency channel since the $\sigma_\alpha$ parameters are analytically marginalized.

\begin{figure}
   \centering
   \begin{tabular}{c}
    \includegraphics[width=0.48\textwidth]{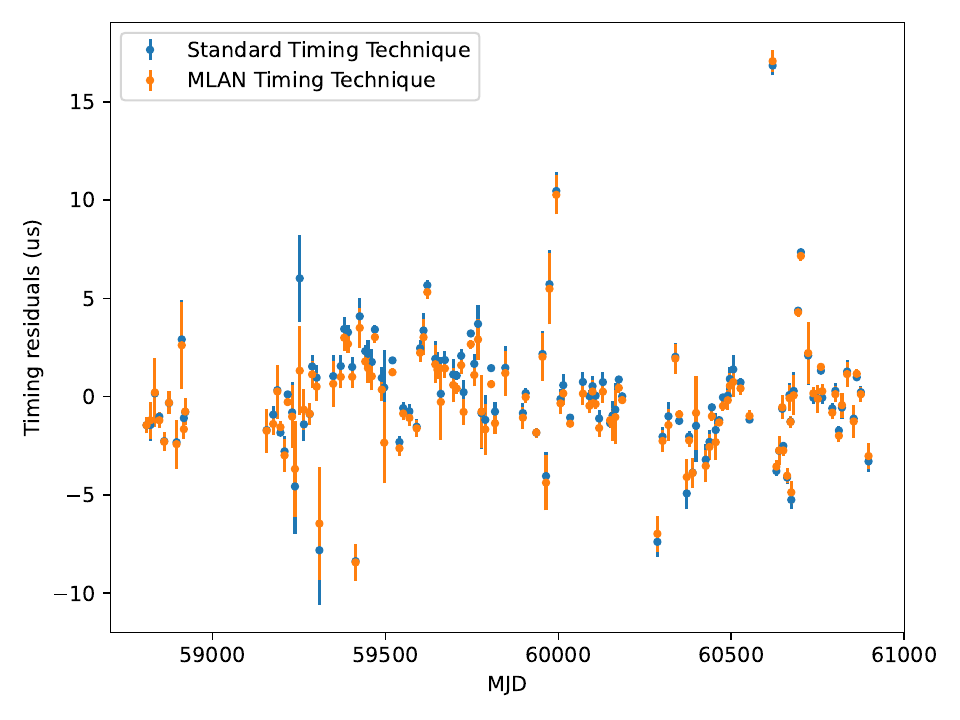}\\
    (a)\\
    \includegraphics[width=0.48\textwidth]{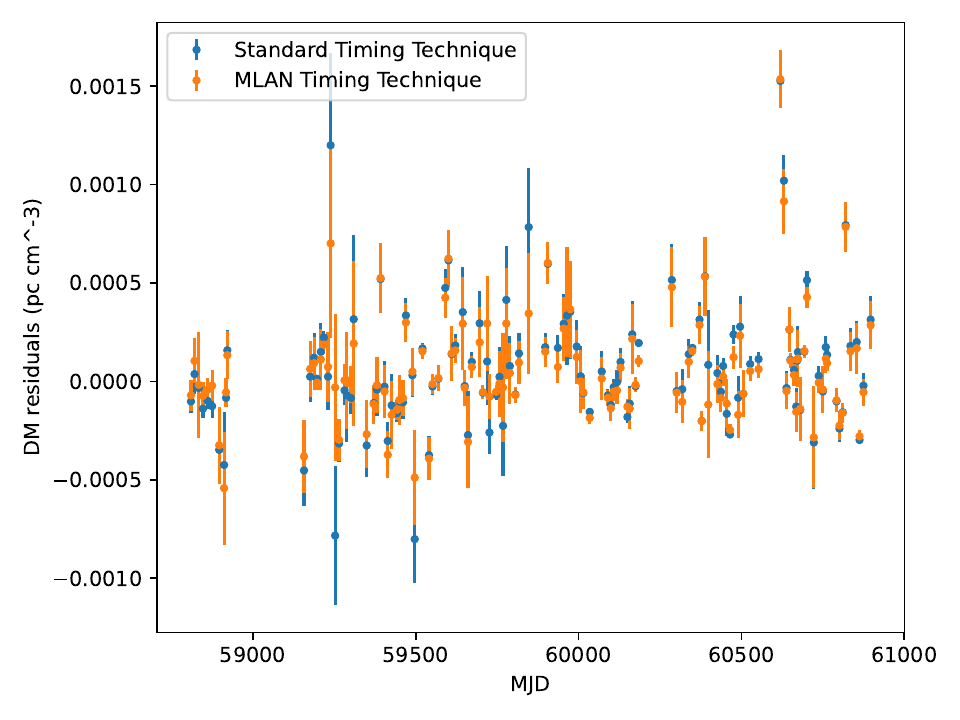} 
    \\(b)
   \end{tabular}
   \caption{The TOA and DM residuals for PSR J2124--3358 derived from preliminary timing of the wideband measurements done using the standard and MLAN methods. The (b) panel of this plot looks similar to the DM measurements shown in Figure \ref{fig:DMtimeseries} because we have only fit a linear function to model the DM during preliminary timing, so the short-timescale DM variations are still present.}
   \label{fig:DM-toa-resid}
\end{figure}

We performed timing analysis on the two sets of TOAs using the \pint{} package \citep{LuoRansom+2021,SusobhananKaplan+2024}.
We start our initial timing analysis using the pulsar ephemeris published in \citet{RanaTarafdar+2025}.
The timing model includes pulsar spin-down parameterized by the frequency F0 and the frequency derivative F1, secular DM evolution parameterized by an average DM and two derivatives (DM1, DM2), and astrometric delays with the sky location (RA, DEC) as free parameters.
These are the same parameters that are fitted in the InPTA Data Release 2 narrowband dataset \citep{RanaTarafdar+2025}.
Proper motion and parallax are not fitted, but fixed to values taken from the European Pulsar Timing Array Data Release 2 \citep{AntoniadisBabak+2023}.
The TOA and DM residuals for standard and MLAN methods are plotted in Figure \ref{fig:DM-toa-resid}.

After the initial timing, we performed single-pulsar noise analysis (SPNA) using a linearized timing model following the method described in \citet{SusobhananVan_Haasteren2025}.
We recall that the Bayes' theorem for SPNA is given by
\begin{align}
    \mathcal{P}[\boldsymbol{\Theta}|\mathfrak{D},\mathfrak{M}] = \mathcal{L}[\mathfrak{D}|\boldsymbol{\Theta},\mathfrak{M}]\;\text{Pr}[\Theta|\mathfrak{M}] / \mathcal{Z}[\mathfrak{D}|\mathfrak{M}]\,,
    \label{eq:post-spna}
\end{align}
where $\mathcal{P}$, $\mathcal{L}$, $\text{Pr}$, and $\mathcal{Z}$ represent the posterior distribution, the likelihood function, the prior distribution, and the Bayesian evidence, respectively, of the SPNA.
Further, $\mathfrak{D}$ represents the wideband TOA and DM measurements, $\mathfrak{M}$ represents the linearized timing and noise model, and $\boldsymbol{\Theta}$ represents the collection of the linearized timing and noise model parameters. 
These are not to be confused with the objects appearing in equations \eqref{eq:post-full} and \eqref{eq:post-marg}, which provide the Bayes theorem in the context of a single wideband measurement.
The SPNA likelihood $\mathcal{L}$ is given by
\begin{equation}
    \mathcal{L} = -\frac{1}{2} \textbf{y}^T \textbf{C}^{-1} \textbf{y} - \frac{1}{2}\ln\det[2\pi \textbf{C}]\, 
\end{equation}
where $\textbf{y}$ is a vector containing the TOA and DM residuals obtained from initial timing, 
$\textbf{C}$ is a covariance matrix usually expressed in a reduced-rank form $\textbf{C} = \textbf{N} + \textbf{M} \boldsymbol{\Phi}\textbf{M}^T$,
$\textbf{N}$ is a diagonal matrix containing modified TOA and DM measurement uncertainties, 
$\textbf{M}$ is a design matrix representing the linearized timing and noise model, 
and $\boldsymbol{\Phi}$ is a prior covariance matrix for the amplitudes associated with the signals represented in $\textbf{M}$.

We implemented the posterior distribution given in equation \eqref{eq:post-spna} with the help of \pint{}.
We model inaccuracies in TOA measurement uncertainties and pulse jitter using an EFAC (error factor) and an EQUAD (error added in quadrature) \citep{KikunagaHisano+2024}. 
Similarly, inaccuracies in DM measurement uncertainties are modeled using a DMEFAC.
We model stochastic DM variations and the achromatic red noise (due to rotational irregularities and gravitational wave background; \citet{AgazieAntoniadis+2024}) as Fourier series Gaussian processes with power-law spectra.  
We use 50 linearly spaced frequency bins each for these Gaussian process models with a fundamental frequency $f_1=T_\text{span}^{-1}$ where $T_\text{span}=2087.5$ days.
Following \cite{Van_HaasterenVallisneri2014b}, we also include four logarithmically spaced frequency bins below $f_1$ to better capture the lower frequency components of the spin noise and DM noise (at $f_1/2$, $f_1/4$, $f_1/8$, and $f_1/16$).
We perform parameter estimation on the wideband measurements using the above-mentioned model with the help of the \texttt{nautilus} package, which implements a neural network-boosted importance nested sampling algorithm \citep{nautilus}.
We perform this analysis on wideband measurements obtained using both the standard and MLAN methods.

\begin{figure*}
    \centering
    \includegraphics[scale=1.1, width=\linewidth]{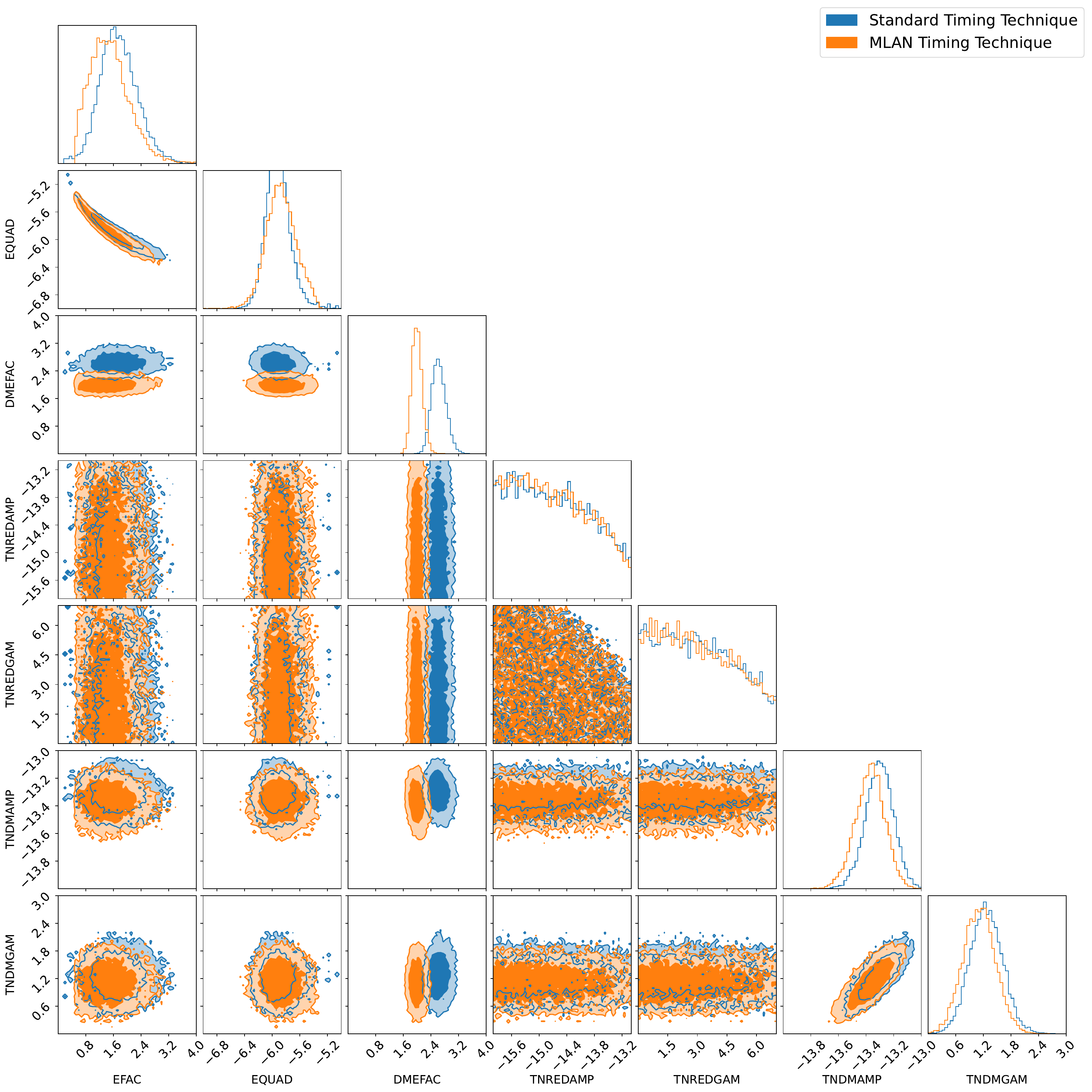}
    \caption{Corner plot showing the SPNA posterior distribution for PSR J2124-3358.
    The results for wideband measurements made using the standard and MLAN methods are displayed in different colors.
    The EQUAD and achromatic red noise parameter estimates are consistent between the standard and MLAN methods.
    On the other hand, the EFAC, DMEFAC, and DM noise estimates are lower for the MLAN method.
    This indicates that the standard method underestimates the measurement uncertainties, whereas the MLAN method provides more realistic measurement uncertainty estimates.}
    \label{fig:noisecorner}
\end{figure*}

The posterior distributions obtained from this exercise are plotted in Figure \ref{fig:noisecorner}.
We find that the DMEFAC estimated from the MLAN measurements is lower than that estimated from the standard measurements.
Similarly, the EFAC and the DM noise amplitude are slightly lower in the case of MLAN.
On the other hand, EQUAD estimates are consistent between the standard and MLAN measurements, although the posterior distributions are not identical.
Finally, achromatic red noise is not detected in this pulsar, and the posterior distributions are visually identical for both standard and MLAN measurements.
These results are consistent with the standard method underestimating measurement uncertainties as seen in Figure \ref{fig:errcomp}.
i.e., the MLAN method provides more realistic measurement uncertainty estimates than the standard method.

\begin{figure}
   \centering
   \includegraphics[width=0.48\textwidth]{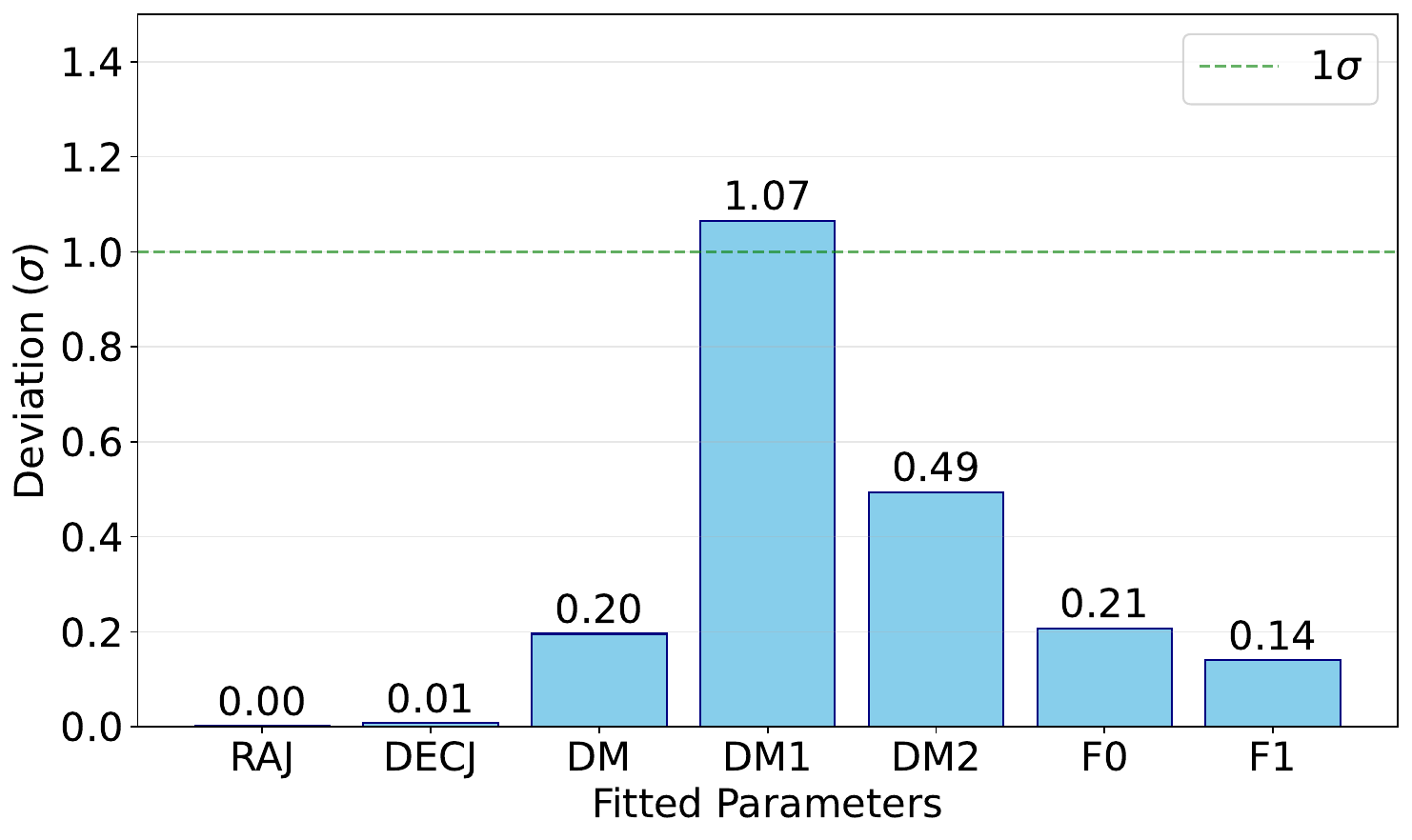}
   \caption{Deviation of timing model parameters estimated from the standard dataset and the MLAN dataset. The parameter estimates are obtained after applying the median noise parameters obtained from SPNA.}
   \label{fig:fitparams_deviation}
\end{figure}

{A comparison of the timing model parameters estimated from the standard and MLAN datasets using \pint{} after applying the median noise parameters obtained from SPNA is shown in Figure \ref{fig:fitparams_deviation}. 
We find that all estimated parameters, except DM1, are consistent within the 1$\sigma$ uncertainties.}

\section{Summary and Discussion}
\label{sec:summary}

We derived a new method for performing wideband TOA and DM measurements from frequency-resolved integrated pulse profiles of a pulsar following a Bayesian approach.
Our method, compared to the widely used wideband measurement method of \citet{PennucciDemorestRansom2014}, features a more rigorous treatment of noise in each frequency channel of the profile, leading to more realistic TOA and DM measurement uncertainty estimates.
Our method, similar to \citet{PennucciDemorestRansom2014}, allows one to make \textit{independent} TOA and DM measurements by choosing an appropriate fiducial observing frequency.
We demonstrated our method using a simulated profile and showed that (a) it accurately recovers the injected parameter values, and (b) the fiducial observing frequency selection procedure is effective in removing the TOA-DM measurement covariance.
We applied our method to the InPTA data of PSR J2124--3318, and compared its performance against the method of \citet{PennucciDemorestRansom2014}.
We showed that the TOA and DM measurement uncertainty estimates obtained using our method are indeed more realistic than those obtained using \citet{PennucciDemorestRansom2014}.
This is crucial for achieving a robust detection of nanohertz gravitational waves using the wideband timing method, since imprecise white noise estimates can lead to biases in pulsar timing array inference \citep{AntoniadisArumugam+2024}.

In the coming years and decades, as the PTA datasets grow, the wideband technique will be crucial for ensuring that the computational cost of performing PTA analysis remains manageable.
This work is a crucial stepping stone towards ensuring the reliability of the wideband timing method, so that it can be applied to high-precision timing experiments like pulsar timing arrays.
We plan to apply our method to the full InPTA dataset in the near future as part of the second InPTA wideband data release.
{However, we note that several major challenges remain, including (1) the rigorous handling of pulse broadening due to interstellar scattering \citep{Sutton1971}, and (2) the handling of mode changes \citep{Backer1970} and other types of profile variability events \citep[e.g.][]{SinghaSurnis+2021}.}
{In addition, we found a few epochs that were severely affected by interstellar scintillation, such that there was very low S/N in some parts of the band, but very high S/N in others.
In such epochs, we found that both the MLAN and standard methods produced imperfect fits to the observed portrait, and the TOAs and DMs estimated by the two methods showed significant differences.
We are currently working on tackling these challenges.}

\section*{Acknowledgments}
We thank the staff of the GMRT who made our observations possible. 
GMRT is run by the National Centre for Radio Astrophysics of the Tata Institute of Fundamental Research.
SJD is supported by IBS under the project code IBS-R018-D1.
SD is supported by SERB grant MTR/2023/000384.
BCJ acknowledges the support from  Raja Ramanna Chair fellowship of the Department of Atomic Energy, Government of India (RRC – Track I Grant 3/3401 Atomic Energy Research 00 004 Research and Development 27 02 31 1002//2/2023/RRC/R\&D-II/13886 and 1002/2/2023/RRC/R\&D-II/14369).
KR is supported by UGC JRF fellowship.
AdS is supported by UGC JRF fellowship
HT is supported by DST INSPIRE Fellowship, INSPIRE code IF210656.
KT is partially supported by JSPS KAKENHI Grant Numbers 20H00180, 21H01130, 21H04467, and 24H01813, and Bilateral Joint Research Projects of JSPS (120237710).
KV is supported by CSIR JRF Fellowship.
ZZ is supported by the Prime Minister’s Research Fellows scheme, Ref. No. TF/PMRF-22-7307.
AS thanks Rutger van Haasteren for fruitful discussions.

\section*{Software}
\texttt{PulsePortraiture} \citep{PennucciDemorestRansom2014,Pennucci2019}, 
\texttt{PINT} \citep{LuoRansom+2021,SusobhananKaplan+2024},
\texttt{nestle} \citep{Barbary2021},
\texttt{nautilus} \citep{nautilus},
\texttt{corner} \citep{Foreman-Mackey2016},
\texttt{numpy} \citep{HarrisMillman+2020},
\texttt{scipy} \citep{VirtanenGommers+2020},
\texttt{matplotlib} \citep{Hunter2007}

\section*{Data Availability}
The wideband TOAs and the ephemeris for PSR J2124--3358 are included as supplementary material.



\bibliographystyle{mnras}
\bibliography{wideband-revisit} 




\appendix

\section{Derivation of the amplitude and noise-marginalized likelihood for wideband measurements}
\label{sec:loglambda}

We begin by recalling the posterior distribution given in equation \ref{eq:post-full}:
\begin{align}
\mathfrak{p}[\varphi_0,D,\boldsymbol{a},\boldsymbol{\sigma}|P,T] =& L[P|\varphi_0,D,\boldsymbol{a},\boldsymbol{\sigma};T] \nonumber\\&\Pi[\varphi_0]\,\Pi[D]\, \Pi[\boldsymbol{a}]\,\Pi[\boldsymbol{\sigma}] / Z[P|T]\,,
\nonumber
\end{align}
where the likelihood function is given by equation \ref{eq:lnlike}:
\begin{align}
\ln L &=-\sum_{\alpha=1}^{N_{\text{chan}}}\left[\sigma_{\alpha}^{-2}\left(U_{\alpha}+a_{\alpha}^{2}V_{\alpha}-2a_{\alpha}W_{\alpha}\right)+
\frac{N_{\text{bin}}}{2}\ln[\pi\sigma_{\alpha}^2]\right]\,.
\nonumber
\end{align}
Assuming an improper uniform prior $\Pi[a_\alpha]\propto 1$ and marginalizing, we obtain 
\begin{align}
\mathfrak{p}[\varphi_{0},D,\boldsymbol{\sigma}|P,T]&\propto\Pi[\varphi_{0}]\,\Pi[D]\,\Pi[\boldsymbol{\sigma}]
\int d^{N_{\text{chan}}}\boldsymbol{a}\;L[P|\varphi_{0},D,\boldsymbol{a},\boldsymbol{\sigma}]\nonumber
\\&=\Pi[\varphi_{0}]\,\Pi[D]\,\Pi[\boldsymbol{\sigma}]L[P|\varphi_{0},D,\boldsymbol{\sigma}]\,,
\end{align}
where the amplitude-marginalized likelihood can be obtained by completing the square in equation \eqref{eq:lnlike} for $a_\alpha$ and evaluating the Gaussian integral.
This gives
\begin{align}
L[P|\varphi_{0},D,\sigma_{\alpha}]
&=\prod_{\alpha=1}^{N_{\text{chan}}}\frac{1}{\left(\sqrt{\pi}\sigma_{\alpha}\right)^{N_{\text{bin}}-1}\sqrt{V_{\alpha}}}\nonumber\\
&\qquad\times\exp\left\{ -\sigma_{\alpha}^{-2}\left(U_{\alpha}-V_{\alpha}^{-1}W_{\alpha}^{2}\right)\right\} \,.
\end{align}
$\mathfrak{p}[\varphi_{0},D,\boldsymbol{\sigma}|P,T]$ can be analytically marginalized over $\boldsymbol{\sigma}$ by assuming an improper log-uniform prior such that $\Pi[\sigma_\alpha]\propto \sigma_\alpha^{-1}$; i.e., 
\begin{align}
\mathfrak{p}[\varphi_{0},D|P,T]&\propto\Pi[\varphi_{0}]\,\Pi[D]\,
\int d^{N_{\text{chan}}}\boldsymbol{\sigma}\;L[P|\varphi_{0},D,\boldsymbol{a},\boldsymbol{\sigma}]\Pi[\boldsymbol{\sigma}]\nonumber\\&=\Pi[\varphi_{0}]\,\Pi[D]\,L[P|\varphi_{0},D]\,,
\end{align}
where the marginalized likelihood can be obtained with the help of the inverse-gamma distribution to be
\begin{align}
    L[P|\varphi_{0},D]	=\prod_{\alpha=1}^{N_{\text{chan}}}\frac{\Gamma\left[\frac{N_{\text{bin}}-1}{2}\right]}{2\sqrt{\pi^{N_{\text{bin}}-1}V_{\alpha}}}\left(U_{\alpha}-V_{\alpha}^{-1}W_{\alpha}^{2}\right)^{\frac{1-N_{\text{bin}}}{2}}\,.
\end{align}
Dropping the factors that are independent of the parameters of interest $\phi_0$ and $D$, we can write the marginalized log-likelihood as
\begin{align}
    \ln \Lambda =\frac{-(N_{\text{bin}}-1)}{2}\sum_{\alpha}\ln\left(U_{\alpha}-\frac{W_{\alpha}^{2}}{V_{\alpha}}\right)\,,
\end{align}
which is a function of only the parameters of interest, $\varphi_0$ and $D$, as well as $\nu_\text{ref}$.

\section{Derivation of the amplitude and noise-maximized likelihood for wideband measurements}
\label{sec:maxlike}
Following \citet{PennucciDemorestRansom2014}, we maximize equation \eqref{eq:lnlike} with respect to the $a_\alpha$ parameters, and and this yields
\begin{equation}
    \hat{a}_\alpha = W_\alpha / V_\alpha\,.
\end{equation}
Substituting, we obtain
\begin{equation}
    \ln L \Big|_{\hat{a}_\alpha}={-}\sum_{\alpha=1}^{N_{\text{chan}}}\left[\sigma_{\alpha}^{-2}\left(U_{\alpha}-\frac{W_{\alpha}^{2}}{V_{\alpha}}\right)+\frac{N_{\text{bin}}}{2}\ln\sigma_{\alpha}^2\right]\,.
    \label{eq:lnlike_ahat}
\end{equation}
We now maximize this expression over the $\sigma_\alpha$ parameters:
\begin{equation}
\hat{\sigma}_{\alpha}^{2}=2N_{\text{bin}}^{-1}\left(U_{\alpha}-\frac{W_{\alpha}^{2}}{V_{\alpha}}\right)\,.
\end{equation}
Substituting this in equation \eqref{eq:lnlike_ahat} and discarding additive constant terms, we obtain
\begin{equation}
    \ln\hat{L}	=\ln L\Big|_{\hat{a}_{\alpha},\hat{\sigma}_{\alpha}}=\frac{-N_{\text{bin}}}{2}\sum_{\alpha}\ln\left(U_{\alpha}-\frac{W_{\alpha}^{2}}{V_{\alpha}}\right)\,,
    \label{eq:loglhat}
\end{equation}

\section{Derivation of the fiducial frequency and the measurement uncertainties}
\label{sec:nurefbar}

When the signal-to-noise ratio in the observed portrait $P$ is large, we expect the distribution of $\varphi_0$ and $D$ to be approximately Gaussian.
Under this assumption, the measurement covariance matrix $\boldsymbol\Xi$ of $\varphi_0$ and $D$ can be approximated as $\boldsymbol{\Xi} = -\textbf{H}^{-1}$,
where 
\begin{equation}
    \textbf{H}= \begin{bmatrix}\frac{\partial^{2}\ln\Lambda}{\partial\varphi_{0}^{2}} & \frac{\partial^{2}\ln\Lambda}{\partial\varphi_{0}\partial D}\\
\frac{\partial^{2}\ln\Lambda}{\partial\varphi_{0}\partial D} & \frac{\partial^{2}\ln\Lambda}{\partial D^{2}}\,.
\end{bmatrix}
\end{equation}
is the Hessian of $\ln \Lambda$.

The measurement covariance can be made to vanish by choosing a value of $\nu_\text{ref}$ such that $\frac{\partial^{2}\ln\Lambda}{\partial\varphi_{0}\partial D}=0$.
We begin by differentiating equation \eqref{eq:varphi} by $\varphi_0$ and $D$:
\begin{subequations}
\begin{align}
   \frac{\partial\varphi_{\alpha}}{\partial\varphi_{0}}&=1\,,\\
   \frac{\partial\varphi_{\alpha}}{\partial D}&=\kappa\bar{F}\left(\nu_\alpha^{-2}-\nu_{\text{ref}}^{-2}\right)\,.
\end{align}    
\end{subequations}
Differentiating equation \eqref{eq:loglhat}, we obtain
\begin{subequations}
\begin{align}
   \frac{\partial\log\Lambda}{\partial\varphi_{\alpha}}=&(N_{\text{bin}}-1)\frac{W_{\alpha}W_{\alpha}'}{\left(U_{\alpha}V_{\alpha}-W_{\alpha}^{2}\right)}\,,\\
   \frac{\partial^{2}\log\Lambda}{\partial\varphi_{\alpha}\partial\varphi_{\beta}}=&\delta_{\alpha\beta}\left(N_{\text{bin}}-1\right)\\&\times\frac{\left(W_{\alpha}'{}^{2}\left(U_{\alpha}V_{\alpha}+W_{\alpha}^{2}\right)+W_{\alpha}W_{\alpha}''\left(U_{\alpha}V_{\alpha}-W_{\alpha}^{2}\right)\right)}{\left(U_{\alpha}V_{\alpha}-W_{\alpha}^{2}\right)^{2}}\,,
\end{align}
\end{subequations}
where, from equation \eqref{eq:W}, we can derive
\begin{subequations}
\begin{align}
    W_{\alpha}'=\frac{\partial W_{\alpha}}{\partial\varphi_{\alpha}}&=-2\pi\sum_{k=1}^{N_{\text{bin}}/2}k\Im\left[\tilde{P}_{\alpha k}\tilde{T}_{\alpha k}^{*}\exp[2\pi ik\varphi_{\alpha}]\right]\,,\\
    W_{\alpha}''=\frac{\partial^{2}W_{\alpha}}{\partial\varphi_{\alpha}^{2}}&=-4\pi^{2}\sum_{k=1}^{N_{\text{bin}}/2}k^{2}\Re\left[\tilde{P}_{\alpha k}\tilde{T}_{\alpha k}^{*}\exp[2\pi ik\varphi_{\alpha}]\right]\,.
\end{align}
\end{subequations}
It is easy to show from the above expressions that
\begin{align}
    \frac{\partial^{2}\ln \Lambda}{\partial D\partial\varphi_{0}}=\kappa\bar{F}\sum_{\alpha}\frac{\partial^{2}\ln \Lambda}{\partial\varphi_{\alpha}^{2}}\left(\nu_{\alpha}^{-2}-\nu_{\text{ref}}^{-2}\right)\,.
\end{align}
Setting this derivative to zero, we obtain
\begin{equation}
\bar\nu_{\text{ref}}^{2}=\frac{\sum_{\alpha}\frac{\partial^{2}\ln \Lambda}{\partial\varphi_{\alpha}^{2}}}{\sum_{\alpha}\frac{\partial^{2}\ln \Lambda}{\partial\varphi_{\alpha}^{2}}\nu_{\alpha}^{-2}}\,.
\label{eq:nurefbar}
\end{equation}
Note that the right-hand side of the above equation is a function of $\varphi_0$, $D$, and $\nu_\text{ref}$.
In practice, we compute $\bar\nu_{\text{ref}}$ by first estimating $\varphi_0$ and $D$ using an arbitrary value of $\nu_\text{ref}$, and then evaluating equation \eqref{eq:nurefbar} using those estimated values.

With this choice of $\nu_\text{ref}$, the measurement variances in $\varphi_0$ and $D$ can be estimated as
\begin{subequations}
\begin{align}
    \sigma_{\varphi_{0}}^{2}=-\left(\frac{\partial^{2}\ln\Lambda}{\partial\varphi_{0}^{2}}\right)^{-1}=&-\left(\sum_{\alpha}\frac{\partial^{2}\ln\Lambda}{\partial\varphi_{\alpha}^{2}}\right)^{-1}\,,\\\sigma_{D}^{2}=-\left(\frac{\partial^{2}\ln\Lambda}{\partial D^{2}}\right)^{-1}=&-\left(\kappa^{2}\bar{F}^{2}\sum_{\alpha}\frac{\partial^{2}\ln\Lambda}{\partial\varphi_{\alpha}^{2}}\left(\nu_{\alpha}^{-2}-\nu_{\text{ref}}^{-2}\right)^{2}\right)^{-1}\,\,,
\end{align}
\end{subequations}
and the TOA variance is 
\begin{equation}
    \sigma^2_{t_\text{arr}} = \bar{F}^{-2}\sigma_{\varphi_{0}}^{2}\,.
\end{equation}

Since $\ln \Lambda$ and $\ln \hat{L}$ have similar functional forms, the $\nu_\text{ref}$ and uncertainty computations given above can be easily repeated for $\ln \hat{L}$.
It is easy to show that the parameter variances estimated from  $\ln \Lambda$ are $\frac{N_\text{bin}}{N_\text{bin}-1}$ times the variances estimated from $\ln \hat{L}$, and hence slightly more conservative.
This difference becomes negligible for large values of $N_\text{bin}$.


\bsp	
\label{lastpage}
\end{document}